\documentstyle[12pt,equation]{article}
\begin{document}
\newcommand{\be}{\begin{equation}}
\newcommand{\ee}{\end{equation}}
\newcommand{\een}{\end{subequations}}
\newcommand{\ben}{\begin{subequations}}
\newcommand{\beq}{\begin{eqalignno}}
\newcommand{\eeq}{\end{eqalignno}}
\newcommand{\lsim}{\begin{array}{c}<\vspace{-0.32cm}\\\sim\end{array}}
\newcommand{\gsim}{\begin{array}{c}>\vspace{-0.32cm}\\ \sim\end{array}}
\pagestyle{empty}
\noindent
OUTP 94-14 P \\
HD-THEP-94-15 \\
hep-th/9408132 \\
August 1994
\vspace{3cm}
\begin{center}
{\bf\Large COLEMAN-WEINBERG PHASE TRANSITION}\\
\medskip
{\bf\Large IN TWO-SCALAR MODELS}\\
\vspace{1cm}
S. Bornholdt$^{\rm a,}$
\footnote{
Address after August $1^{st}$: Institut f\"ur Theoretische
Physik, Universit\"at Kiel, Olshausenstr. 6, 24118 Kiel,
Germany.},
N. Tetradis$^{\rm b}$
and
C. Wetterich$^{\rm a}$\\
\bigskip
a)Institut  f\"ur Theoretische Physik, Universit\"at Heidelberg,\\
Philosophenweg 16, 69120 Heidelberg, Germany\\
b)Theoretical Physics, University of Oxford,\\
1 Keble Road, Oxford OX1 3NP, U.K.\\
\vspace{1cm}
\abstract{
We explore the Coleman-Weinberg phase transition in regions outside
the validity of perturbation theory. For this purpose we
study a Euclidean field theory with two scalars and discrete symmetry
in four dimensions.
The phase diagram is established by a numerical solution of a
suitable truncation of exact non-perturbative flow equations.
We find regions in parameter space where the phase transition
(in dependence on the mass term) is of the second or the
first order, separated by a triple point.
Our quantitative results for the first order phase
transition compare well to the
standard perturbative Coleman-Weinberg
calculation of the effective potential.}
\end{center}

\clearpage

\setlength{\baselineskip}{15pt}
\setlength{\textwidth}{16cm}
\pagestyle{plain}
\setcounter{page}{1}

\newpage

\setcounter{equation}{0}
\renewcommand{\theequation}{\arabic{equation}}

During the last two decades, the perturbative
Coleman-Weinberg approach \cite{CW} to
the calculation of the effective potential by a loop expansion has been
widely used in studying
field theories that exhibit spontaneous symmetry breaking.
Within the standard model and
for low masses
of the Higgs boson this approach predicts a weakly first order phase
transition in dependence on the mass parameter. This feature has been
used to derive a lower bound on the mass of the Higgs boson \cite{LOWB}.
A generalization of the Coleman-Weinberg calculation for
non-vanishing temperature suggests also a first order phase transition
in dependence on the temperature, at least if the mass of the Higgs Boson
is sufficiently small \cite{PTSM}. The observation that the baryon
asymmetry may have been created during the electroweak phase transition
in the very early universe \cite{BAU} considerably renewed the interest
in this subject \cite{2a}.
Indeed, this hypothesis received much encouragement from the
fact that the loop expansion for the
potential predicts a first order phase
transition with periods outside thermodynamic equilibrium,
which is one of the three main ingredients for
the generation of net baryon number \cite{SAKH}.
When the experimental bounds for the
Higgs mass \cite{BOUND} began to diverge from the theoretical expectations
\cite{LIM}, two-Higgs-doublet extensions of the standard model entered
the scene. They provide  additional CP violation (which is another requirement
for baryogenesis). Also perturbative calculations of the effective
potential suggest that the phase transition in these models can be
sufficiently strongly first order and nevertheless remain compatible
with the experimental mass bound \cite{AH}. However,
it is not clear whether these perturbative methods give the right
picture of the phase transition, since the same methods fail to
give a correct description of the second order high temperature
phase transition \cite{TW} in models with only one scalar.
A complete description of
the phase transitions in two-scalar models is still lacking.

Another wide range of application of effective potentials concerns
inflationary cosmology.
In the model proposed originally by Guth \cite{GUTH}
as well as in later models \cite{MODELS},
an effective potential of the Coleman-Weinberg type has been used to
allow for a phase of exponential expansion of the early universe.
These models
often do not specify what the underlying theory
should be, but they nevertheless assume that a general field theoretical
framework exists which produces this type of potential.

The high temperature phase transition can be understood in terms
of the temperature dependent effective potential
$U(\varphi,T)$ for the scalar field $\varphi$.
The crucial feature of a first order transition
is the existence of two local minima at $\varphi=0$ and
$\varphi \not= 0$, separated by a barrier, in a certain
temperature range. Similarly, if the zero temperature
theory exhibits this feature for a certain range of the mass
parameter one speaks about a first order transition in
dependence on the mass parameter. By continuity in $T$
for $T=0$,
the two phenomena are closely related.
The renormalization group improved loop expansion is believed to
work well near the absolute minimum of the effective potential.
Already in the original paper \cite{CW}, however, it was pointed
out that the description of possible additional relative
minima is not within the range of applicability of perturbation
theory. In the phase with spontaneous symmetry breaking the
validity of the one-loop results for the form of the effective
potential at the origin ($\varphi=0$)
has not been established so far. However, a reliable description
of both the absolute and the relative minimum is crucial
for a firm establishment of a first order phase transition.

In this letter we perform a non-perturbative analysis
which is capable of dealing with the above problem.
It confirms the picture arising from
the results of the Coleman-Weinberg loop expansion
for the effective potential in the vicinity of the
phase transition. We employ a new method based on the effective average action
\cite{AA} which is the analogue of the block spin action
\cite{WK} for continuous space. The dependence of the effective average
action $\Gamma_k$ on the average scale $k$ (one averages over a volume
$\sim k^{-d}$) is given by an exact evolution equation \cite{EXAA}
\footnote{See ref. \cite{reneq} for other version of exact
renormalization group equations.}.
In the limit $k\rightarrow 0$,
$\Gamma_k$ becomes the usual effective action (the generating
functional of the 1PI Greens functions).
For $k>0$ the effective average potential $U_k$
(the non-derivative part of $\Gamma_k$)
is not necessarily convex. This permits the study of
several minima of different depth. The effective average potential
becomes convex in the limit $k \rightarrow 0$
\cite{convex}.

We want to solve the flow equation which describes how the
effective average potential changes its shape as $k$ decreases
from some high scale (UV cutoff) to zero. For this purpose we
first employ a truncation of the exact non-perturbative evolution equation
by neglecting wave function renormalization effects and $\varphi^6$
and higher effective vertices. Even in this approximation our approach
goes beyond the usual gap equation for the mass term since it effectively
contains a similar self-coupled equation for the quartic scalar coupling.
It also properly accounts for mass effects in the scale dependence of 1PI
vertices and describes how particles with mass $m$ much larger
than $k$ decouple from the running of the couplings with
varying $k$.
These features are not included in the Coleman-Weinberg one-loop
calculation. The standard perturbative result for the
$\beta$-functions of the quartic scalar couplings is recovered from our
results for small couplings and vanishing masses.
In a second step we improve the truncation by including
the effects of
$\varphi^6$ and $\varphi^8$ couplings.
This permits the investigation of potentials with a rich
variety of shapes. For example, the potential at large $k$ may have
only one minimum (up to the degeneracy due to symmetry),
whereas at lower $k$ new relative minima may appear as a result of
integrating out quantum modes with momenta larger than $k$.
Following the evolution of the different extrema permits the
determination of the nature of the phase transition.
The reliability of our non-perturbative method for second order phase
transitions has recently been
demonstrated by computing
the high temperature phase transition of the O(N)-symmetric scalar theory
and the corresponding critical exponents
to a high precision \cite{TW}.
Here we address a model with a first order phase transition.

We consider a simple scalar field theory with two scalar degrees of freedom
in four dimensions. We limit ourselves to the zero temperature case
and study the phase transition with varying mass parameter.
After discussing the tree level potential we first recall the results
of the Coleman-Weinberg loop expansion.
Then the method of the effective average action is briefly introduced
and the formalism for the model under investigation is given.
We study spontaneous symmetry breaking in a model
of two scalar fields $\varphi_1$, $\varphi_2$
with a classical potential symmetric under
$\varphi_1 \rightarrow - \varphi_1,\;
\varphi_2 \rightarrow - \varphi_2,\;
\varphi_1 \rightarrow  \varphi_2$
\be
V = -\mu^2(\rho_1+\rho_2) + \frac{\lambda}{2}(\rho_1^2+\rho_2^2)
  + g \; \rho_1 \rho_2,
\label{POT1} \ee
where
$\rho_1 = \frac{1}{2} \varphi_1^2$ and
$\rho_2 = \frac{1}{2} \varphi_2^2$.
This potential is bounded from below provided $g > - \lambda, \, \lambda > 0$.
It is also convenient to use the parametrization
\be
V = -\mu^2(\rho_1+\rho_2) + \lambda \; \left[
\frac{1}{2}(\rho_1+\rho_2)^2 +x \; \rho_1 \rho_2 \right]
\label{POT}
\ee
with $x = (g-\lambda)/\lambda$ measuring the deviation from the case with
$O(2)$ symmetry discussed earlier with the method of the effective
average action
\cite{AA,TW}.
The structure of
this potential is described by three different phases.
For $\mu^2 < 0$ the classical theory is in the
symmetric phase with the minimum lying at the origin (S phase).
If  $\mu^2 > 0$, one or both of the scalar fields develop
a vacuum expectation value. One finds
the minimum on the axes
for $x>0$ (AX phase) or
between the axes for $x<0$ (M phase).
For $x=0$ the potential exhibits an additional O(2) symmetry.
Due to the symmetries of the potential, the minima on the axes occur
symmetrically at
$(\rho_1 = \mu^2/\lambda \; , \; \rho_2=0)$ and
$(\rho_1=0 \; , \; \rho_2 = \mu^2/\lambda)$.
In the M phase, the minimum is always
located between the axes at
$\rho_1 = \rho_2 = \mu^2/(g + \lambda)$.

The renormalization group improved one-loop potential can be
parametrized in terms of scale dependent
couplings $\lambda$ and $g$.
An important ingredient is the coupling $g$ which
couples the two scalar fields. In a study of the scale dependence of
$\lambda$ this coupling appears in the renormalization
group equation for $\lambda$ (and vice versa).
Two interesting scenarios might be induced,
which seem to indicate first order phase transitions.
First, a term $\sim g^2$
appears in the $\beta$-function for $\lambda$, so that $\lambda$,
instead of running
to zero for $k \rightarrow 0$ as in the O(N) symmetric $\varphi^4$-theory,
may reach zero
during a finite renormalization interval and
then be driven to negative values.
For $\mu^2 > 0$ (AX phase) this would correspond to turning the minimum
of the
effective potential into a saddlepoint. This situation is very similar to the
scale dependence of the quartic scalar coupling in the Abelian Higgs model
originally studied by Coleman and Weinberg \cite{CW} (with $g$ playing the
role of $e^2$) and is generally believed to indicate a first order phase
transition.
A second interesting case occurs for negative couplings $g$.
When $g$ becomes smaller than $-\lambda$, again the minimum flips over.
Let us see more quantitatively what standard perturbation theory tells us.
The one-loop renormalization group equation for the quartic coupling
$\lambda$ reads
\be
\frac{d \lambda}{d t} = \frac{1}{16\pi^2} \; (9\lambda^2 + g^2),
\label{DL}
\ee
with a logarithmic renormalization scale $t= \ln k$.
Indeed, If $g^2$ is large enough, $\lambda$ runs to zero within a finite
interval of $t$. Since the evolution of the mixed coupling $g$ turns out to
be
\be
\frac{d g }{d t} = \frac{1}{16\pi^2} \; \left( 6\lambda g + 4g^2 \right),
\label{DG}
\ee
this interesting feature always shows up for a sufficiently large starting
value $g_0/\lambda_0$.
In the case of negative couplings $g$ we are interested
in what happens for $x$ approaching $-2$ or $g \simeq -\lambda$.
Near $x\simeq-2$ one finds
\be
\frac{d x}{d t} \simeq \frac{\lambda}{2 \pi^2}
\label{DX}
\ee
which, for large enough $\lambda$ and initial value of $x$ close to $-2$,
leads to values of $x<-2$ thus flipping the minimum into a saddlepoint!
In the same region one has
\be
\frac{d \lambda}{d t} \simeq \frac{5 \lambda^2}{8\pi^2},
\label{lala} \ee
ensuring that there is always a natural parameter range where this phenomenon
occurs,
indicating again a first order phase transition.

However, some questions remain open.
The above renormalization group equations do not
depend on the masses of the theory and are in this respect
``scale invariant''.
This is a good approximation for mass scales much larger than the
characteristic masses of the theory. For a first order phase
transition
the masses do not vanish even for the critical parameters.
For scales below the critical mass the running of the
dimensionless couplings should stop. For a precise understanding
of the character of the phase transition we want to take these
``threshold effects'' into account. We therefore introduce a scale
$k$
which acts as an infrared cutoff for the quantum fluctuations,
independently of all other possible infrared cutoffs such as
field dependent masses and external momenta.
This replaces the
effective potential $U$ by an ``effective average potential''
$U_k$ in which only field modes with
momenta larger than $k$ are integrated over.

We start with the evolution equation \cite{EXAA} describing the
dependence of the effective average potential $U_k$ on the average scale
$k (t = \ln k)$
in arbitrary dimensions $d$
\be
\frac{\partial U_k}{\partial t}(\rho_1,\rho_2) = v_d\;
\int\limits_0^\infty
dx\; x^{\frac{d}{2}-1} \; \frac{\partial P}{\partial t}
\left\{ \frac{1}{P+M_1^2} + \frac{1}{P+M_2^2} \right\}.
\label{MAST}
\ee
Here the variable $x$ denotes momentum squared
and $v_d^{-1} = 2^{d+1} \pi^\frac{d}{2} \Gamma(\frac{d}{2})$.
The inverse average propagator
\begin{equation}
P = \frac{x}{1-\exp \left(-\frac{x}{k^2} \right)}
\label{PE}
\end{equation}
contains an effective infrared cutoff for the modes with $x<k^2$. The
eigenvalues of the mass matrix
\be
M_{1,2}^2 = \frac{1}{2} (U_1 + U_2) +U_{11}\rho_1 + U_{22}\rho_2
\pm \sqrt{ \left[\frac{1}{2} (U_1-U_2)
+ U_{11}\rho_1 - U_{22}\rho_2 \right]^2 + 4 U_{12}^2 \rho_1 \rho_2 }
\label{MS}
\ee
are determined by the partial derivatives of the effective average potential
$U_1 \equiv \partial U_k/\partial \rho_1$,
$U_{12} \equiv \partial^2 U_k/\partial \rho_1\partial \rho_2$ etc.
We note the close resemblance of the flow equation (\ref{MAST})
to a renormalization group improved one-loop calculation with
infrared cutoff (for $k=0$ one has $P=x$). Up to neglecting effects
from the wave function renormalization this flow equation is an exact
non-perturbative equation \cite{EXAA,TW}.
The appearance of renormalized and $\rho$-dependent
mass terms instead of bare mass terms turns eq. (\ref{MAST}) into a
self-coupled equation with a structure similar to the gap equation for the
renormalized mass. A gap equation for the masses can indeed be obtained by
computing $\partial M^2_{1,2}/\partial t$ using partial derivatives of
eq. (\ref{MAST}) with respect to $\rho_1$ and $\rho_2$.
Similar self-coupled equations for
the quartic couplings follow from taking second partial
derivatives of eq. (\ref{MAST}).
After performing the momentum integration the evolution equation (\ref{MAST})
becomes a partial differential equation for $U_k(\rho_1,\rho_2,t)$.
This is difficult to solve explicitly and we will first
discuss an approximate
solution based on neglecting partial derivatives of higher than second order,
i.e.\ terms $\sim U_{111}$ or $\varphi^6$ couplings and so on.
This approximation is too crude to account well for complex structures
of
$U_k$, as for example several inequivalent local minima (see below).
Nevertheless it is a valid approximation as long as $k$ is much
larger than the physical masses. Before turning to less
restrictive truncations we use the quartic potential here as
an illustration of our method.
The evolution of the potential is then described by a set of three
renormalized parameters which depend on the average scale $k$.
We choose $\lambda = U_{11}$, $x = U_{12}/U_{11} -1$ where partial
derivatives are evaluated at the minimum, and
a third parameter which is either a mass $m^2 = U_1$ (S phase) or the
distance $\rho_0$ of the minimum from the origin, i.e.
$\rho_0 = (\rho_1)_{\rm min}$ (AX phase) or
$\rho_0 = (\rho_1)_{\rm min}+(\rho_2)_{\rm min}$ (M phase).
At the ultraviolet cutoff $\Lambda$, the physics is described
by the classical potential $V$, while at some lower scale $k<\Lambda$
by the effective average potential $U_k$ which is obtained from
eq. (\ref{MAST}) with the initial condition $U_\Lambda = V$.
The effective average potential interpolates
between the classical potential
($k=\Lambda$) and the effective potential ($k=0$) \cite{EXAA}.
A solution of the evolution equation for $k\rightarrow 0$ then leads to the
effective 1PI vertices at zero momentum. The effective average potential does
not in general stay a quartic potential but receives higher order
terms from the renormalization group flow. Near a unique
minimum, however,
it can be taken to a good approximation to be of the same form as
in eq. (\ref{POT})
with renormalized parameters $\mu^2$, $\lambda$, and $x$. In this case
our truncation of higher order couplings (like $\varphi^6$ terms)
should be a valid approximation.
The evolution of the renormalized parameters with changing scale $k$
is derived from eq. (\ref{MAST}) by partial differentiation with respect
to $\rho_1$, $\rho_2$
\footnote{
Note that in the AX phase the evolution equations for $\lambda = U_{11}$
would differ from the one defined as  $\lambda = U_{22}$ by higher order
terms in powers of $\lambda$ and $x$. This is due to the choice of a
particular truncation of the effective
average potential $U_k$ which we fix by
defining $\lambda$ always through partial derivatives in the direction
of the minimum as viewed from the origin.
}.

We next present the evolution equations for the
three relevant couplings for the
S regime, where the partial derivatives defining the running parameters
are evaluated at the minimum which lies at the origin
$ (\rho_{10} = \rho_{20} =0) $. This is always the case for
$m^2 = -\mu^2 > 0$.
One obtains
\beq
\frac{d m^2}{d t} =& -2v_d k^{d-2} (4+x) \; \lambda \;
l_1^d \, s_1^d \left( \frac{m^2}{k^2} \right)
\label{DMS}
\\
\frac{d \lambda}{d t} =& 2v_d k^{d-4} (10+2x+x^2) \;
\lambda^2 \; l_2^d \, s_2^d \left( \frac{m^2}{k^2} \right)
\label{DLS}
\\
\frac{d x}{d t} =& -2v_d k^{d-4} (x+1)x(x-2) \;
\lambda \; l_2^d \, s_2^d \left( \frac{m^2}{k^2} \right),
\label{DXS}
\eeq
where
$l_1^d$ and $l_2^d$ are constants of order one defined by
\be
l_n^d = \frac{n}{2} k^{2n-d} \int\limits_0^\infty
dx \; x^{\frac{d}{2}-1} \frac{\partial P}{\partial t} P^{-n-1}.
\label{ldn}
\ee
The ``threshold functions''
$s_{n}^d \left( \frac{w}{k^2} \right)$ are given by
\be
s_n^d \left( \frac{w}{k^2} \right) =
\frac{n}{2} (l_n^d)^{-1} k^{2n-d} \int\limits_0^\infty
dx \; x^{\frac{d}{2}-1} \frac{\partial P}{\partial t} (P + w)^{-n-1}
\label{sdn}
\ee
and depend on the dimensionless ratio
$w/k^2$. They approach unity for vanishing argument,
vanish for large arguments and describe the decoupling of particles with
mass greater than $k$. More details can be found in refs. \cite{AA,TW}.
Another useful equation, though redundant, is the running of the
coupling $g$
\be
\frac{d g}{d t} = 2v_d k^{d-4} (10+14x+4x^2) \;
\lambda^2 \; l_2^d \, s_2^d \left( \frac{m^2}{k^2} \right)
\label{DGS}
\ee
with $g=(1+x)\lambda$. We observe that for $d=4$ and $m^2 = 0$ we
recover from eqs. (\ref{DLS}) and (\ref{DGS}) the standard one-loop
$\beta$-functions of eqs. (\ref{DL}) and (\ref{DG})
(making use of $v_4 = 1/32\pi^2$
and $l_2^4 = 1$).

The only non-perturbative content in the approximation of eqs.
(\ref{DLS}) and (\ref{DGS}) arises through the threshold functions
$s_2^d$ which describe in a natural way that particles with $m^2 \gg k^2$
should not influence the variation of the couplings with an infrared scale
$k$.
The presence of mass  thresholds in the $\beta$-functions, though physically
very reasonable, is not seen in many versions of the renormalization group
equations. In our approach it arises naturally from eq. (\ref{MAST}),
together with an additional equation (\ref{DMS}) for the scale dependence
of the mass term.
The ``quadratic renormalization'' of the mass term given by eq. (\ref{DMS})
incorporates in the renormalization group framework
the physics related to the ``quadratic
divergences''. Although known in
practice since a long time in the Wilson approach to the renormalization group
(for example in lattice studies) this equation is often missing in the
framework of perturbative renormalization group equations. The reason is
simply that if no mass scale other than $m^2$ is present in a given
formulation of the renormalized potential, ratios like $m^2/k^2$
cannot be formed and such approaches are necessarily blind to
mass thresholds.

In the AX regime ($x>0$) we choose the minimum at
$ ( \rho_{10} = \rho_0 > 0 \; , \; \rho_{20} = 0 ) $.
Here $\rho_0$ is determined by $U_1(\rho_0,0) = 0$ and one has
$U_2(\rho_0,0) = (g-\lambda)\rho_0 = x\lambda\rho_0$.
The evolution equations now refer to the couplings defined at
$(\rho_0,0)$
\beq
\frac{d \rho_{10}}{d t} =& 2v_d k^{d-2} \; l_1^d \; \left[ 3 \;
s_1^d(2\lambda\kappa) + (1+x) \; s_1^d(x\lambda\kappa) \right]
\label{DRAX}
\\
\frac{d \lambda}{d t} =& 2v_d k^{d-4} \; l_2^d \; \lambda^2 \left [9 \;
s_2^d(2\lambda\kappa) + (1+x)^2 \; s_2^d(x\lambda\kappa) \right]
\label{DLAX}
\\
\frac{d x}{d t} =& 2v_d k^{d-4}\; l_2^d \; \lambda \;
\frac{x(1+x)}{1-\frac{x}{2}}
\; \left[ 9 \, s_2^d(2\lambda\kappa)
+ \left( \frac{x^2}{2} -2x -7 \right) \; s_2^d(x\lambda\kappa) \right]
\label{DXAX}
\\
\frac{d g}{d t} =& 2v_d k^{d-4} \; l_2^d \; \lambda^2 \left[\frac{
9 +\frac{27}{2}x + \frac{9}{2}x^2}{1-\frac{x}{2}} \;
s_2^d(2\lambda\kappa)
+(1+x) \; \frac{1-\frac{11}{2}x - 2x^2}{1-\frac{x}{2}} \;
s_2^d(x\lambda\kappa) \right]
\label{DGAX}
\eeq
with $\kappa=\rho_0/k^2$.
If finally $-2 < x < 0$ and both scalars develop a vacuum expectation
value (M regime) we use $\rho_{10} = \rho_{20} = \frac{1}{2} \rho_0$
with $\rho_0$ determined by $U_1(\frac{1}{2}\rho_0,\frac{1}{2}\rho_0) =
U_2(\frac{1}{2}\rho_0,\frac{1}{2}\rho_0) = 0$.
The mass eigenvalues are given by
$M_1^2 = (2+x)\lambda\rho_0 $, $M_2^2 = -x\lambda\rho_0 $
and the evolution equations read
\beq
\frac{d \rho_0}{d t} =& 2v_d k^{d-2} l_1^d \; \left[ 3
s_1^d((2+x)\lambda\kappa) + \frac{2-x}{2+x} \; s_1^d(-x\lambda\kappa) \right]
\label{DRM}
\\
\frac{d \lambda}{d t} =& 2 v_d k^{d-4} \;
l_1^d \; \frac{3 x \lambda}{\kappa} \;
\frac{1+\frac{x}{4}}{1+x} \; \left[ s_1^d((2+x)\lambda\kappa)
- s_1^d(-x\lambda\kappa) \right]
\nonumber \\
+& 2v_d k^{d-4} \; l_2^d \;
\lambda^2 \; \left[ 9 \left( 1+\frac{x}{2} \right)^2 \;
s_2^d((2+x)\lambda\kappa) +
\left( 1-\frac{x}{2} \right)^2 \;
s_2^d(-x\lambda\kappa) \right]
\label{DLM}
\\
\frac{d x}{d t} =& -2v_d k^{d-4} \; l_1^d \; \frac{3}{\kappa} \;
\frac{2+x}{1+x} \; \left( x+\frac{x^2}{4} \right) \;
\left[ s_1^d((2+x)\lambda\kappa) - s_1^d(-x\lambda\kappa) \right]
\nonumber \\
-& 2v_d k^{d-4} \; l_2^d \; x \lambda \;
\left[ 9 \left( 1+\frac{x}{2} \right)^2 \;
s_2^d((2+x)\lambda\kappa)
+ \left( 1-\frac{x}{2} \right)^2 \; s_2^d(-x\lambda\kappa) \right]
\label{DXM}
\\
\frac{d g}{d t} =& -2v_d k^{d-4} \; l_1^d \; \frac{3 x \lambda}{\kappa} \;
\frac{1+\frac{x}{4}}{1+x} \; \left[ s_1^d((2+x)\lambda\kappa)
- s_1^d(-x\lambda\kappa) \right]
\nonumber \\
+& 2v_d k^{d-4} \; l_2^d \;
\lambda^2 \; \left[
9 \left( 1+\frac{x}{2} \right)^2 \; s_2^d((2+x)\lambda\kappa)
+ \left( 1-\frac{x}{2} \right)^2 \; s_2^d(-x\lambda\kappa) \right].
\label{DGM}
\eeq
Again the perturbative one-loop equations of $\lambda$ and $g$ are
recovered for $d=4$ by
putting $s_2^4=1$ in eqs. (\ref{DLAX}), (\ref{DGAX}) or
eqs. (\ref{DLM}), (\ref{DGM}) and non-perturbative effects arise only
through the mass dependence of the threshold functions.

With the three sets of evolution equations we are now prepared to
study the phase transition in the two-scalar model in four dimensions.
Fig. 1 shows the flow of the coupling $x$ and the dimensionless ratio
$\kappa = \rho_0/k^2$ which indicates the position of the minimum of the
potential. Trajectories are shown
with decreasing scale $t=\ln (k/\Lambda)$, according to the sets of
evolution equations given above with initial conditions set at some high
momentum scale $\Lambda$.
The phase diagram of fig. 1 is obtained for $\lambda(\Lambda) = 0.1$.
We observe the three phases S, AX and M separated by (solid) phase
transition lines. The phase
transition between the AX and M phase occurs always
for $x=0$. This is easy to understand since $x=0$ corresponds to an
enhanced O(2) symmetry. This symmetry is preserved if we start
at $\Lambda$ with an O(2) symmetric potential. Since no trajectory can
cross the line $x=0$ the phase diagram actually splits
into two separate
parts for $x>0$ and $x<0$. The phase transition between the AX and M
phases as a function of $x(\Lambda)$ is of the second order and we
observe that trajectories with small $x$ are attracted towards the
transition line.

The phase transition between the S phase and SSB phase (AX or M phase)
does not occur for $\kappa =0$ as in the tree approximation
but rather for positive $\kappa(\Lambda)$.
In the S phase near the phase transition
a typical trajectory starts at $\Lambda$ near the critical line at
$\kappa(\Lambda) >0$.
It stays near the critical line for a certain range of scales, moving
according to the arrows as dictated by the evolution equations in the
AX or M phase. At a certain scale $k_c$ it deviates strongly from the
critical line and $\kappa$ reaches zero at some scale $k_s >0$.
Then the evolution equation of the symmetric regime has to be used,
with boundary condition $m^2(k_s) =0$.
The true phase diagram is, of course, three-dimensional and we only show a
projection. The trajectories also have a component perpendicular to the
projected plane. Since for small $\lambda$ the evolution of $\lambda$
is very slow, the projection on the plane with constant $\lambda$
gives a satisfactory picture.

On the critical line one observes three fixpoints:
The ``Ising fixpoint'' at $x=-1$ corresponds
to two disconnected scalar theories. The ``Heisenberg fixpoint'' at $x=0$
is characterized by
an additional O(2) symmetry, and the ``cubic fixpoint'' is situated near
$x=2$ (compare with eq. (\ref{DXS}))
\footnote{The true fixpoint occurs exactly for $x=2$ and the small deviation
from this value is due to a particularity of our truncation.
For details see ref. \cite{BTW}.}.
The three points are partial fixpoints
and become full fixpoints
only in the limit where the running of $\lambda$ is neglected. (The
only true full fixpoint is the Gaussian fixpoint for $\lambda=0$.)
They become genuine fixpoints in three (or $4-\epsilon$) dimensions
and we have used here their names common in statistical mechanics
\cite{AMIT}. The fixpoints at $x=-1$ and $x=2$ are repulsive
and the O(2) symmetric one at $x=0$ is attractive.

In the region $-1<x<2$ the phase transition is of the second order.
Near-critical
trajectories stay very near the phase transition line before
$\kappa$ deviates either towards zero (S phase) or infinity (SSB phase)
at some scale $k_c$.
Exactly on the phase transition the trajectories stay on the critical line
for infinitely long ``time'', i.e. $k_c \rightarrow 0$.
For $x<-1$ the critical trajectories run towards $x=-2$ where the
quartic polynomial is unstable. We will show that in this region the
phase transition is of the first order.
The point $x=-1$ is a tricritical point separating the first and
second
order part of the phase diagram. Similarly, for $x>2$ the trajectories
run to $x \rightarrow \infty$ ($\lambda \rightarrow 0$).
This part of the phase diagram corresponds again to a first order
transition with tricritical point at $x=2$.
For the establishment of the first order character of the phase
transition the two interesting regions are
$x \simeq -2$ and $x \rightarrow \infty$. They lie on the left and
right edges of the phase diagram in fig. 1.

In the regions $x \simeq -2$ and $x \rightarrow \infty$
($\lambda \simeq 0$) the approximation of the effective average
potential $U_k(\rho_1,\rho_2)$ by a quadratic polynomial
in $\rho_1$ and $\rho_2$ becomes inadequate. Indeed, if one
would insist on solving the corresponding evolution equations
discussed before, one would find in these regions
an asymptotic behaviour of the phase transition line
characterized by fixpoints in
$\kappa^3 \lambda (g + \lambda)$ and
$\kappa^3 \lambda g$
respectively. They simulate a second order transition and are
artefacts of an insufficient truncation.
One has to account for the possibility that the effective average
potential has more than one local minima in these regions.
In this case a description of the
potential in terms of ``local variables'' at the absolute minimum,
i.e. a polynomial expansion around this minimum, may
be insufficient. A more global approach, which permits the
simultaneous existence of more than one local minima, and monitors
their properties as well as the differences in height between them,
seems more appropriate.

In the remaining part we will extend the study  by including higher
derivative couplings as well as by monitoring
the extrema of the potential simultaneously, in order to
to get an accurate
picture of the phase transition.
We restrict ourselves to the region $x \rightarrow \infty$
($\lambda \simeq 0$)
\footnote{The region $x \simeq -2$ can be mapped onto
this region by a field transformation \cite{BTW}.}.
We parametrize the effective
average potential $U_k$ by eight parameters
\begin{eqnarray}
U_k &=& -\mu^2(\rho_1+\rho_2) + \frac{1}{2}\lambda_0(\rho_1^2+\rho_2^2)
+ g_0\rho_1\rho_2 \nonumber \\
&+& \nu_A(\rho_1^2\rho_2 + \rho_1\rho_2^2) + \nu_B(\rho_1^3 + \rho_2^3)
\nonumber \\ & +&
\gamma_A\rho_1^2\rho_2^2 + \gamma_B(\rho_1^3\rho_2 + \rho_1\rho_2^3) +
\gamma_C(\rho_1^4+\rho_2^4).
\label{pot8}
\end{eqnarray}
A convenient set of running parameters in this region is
\beq
m_0^2 &= U_1(0) = U_2(0) = -\mu^2~~~~~~~~
\lambda_0 = U_{11}(0) = U_{22}(0)~~~~~~~~
g_0 = U_{12}(0)  \nonumber \\
&\rho_0~~~ {\rm from}~~~  U_1(\rho_0) =0~~~~~~~~
m^2 = U_2(\rho_0)  \nonumber \\
\lambda_1 &= U_{11}(\rho_0)~~~~~~~~
\lambda_2 = U_{22}(\rho_0)~~~~~~~~
g = U_{12}(\rho_0).
\label{var8}
\eeq
Here we assume
a vacuum expectation value at ($\rho_{10}=\rho_0$,\ $\rho_{20}=0$)
and the partial derivatives of the potential are
evaluated at $\rho_1=0$ and $\rho_1=\rho_0$. The first three parameters
describe the potential at the origin while the last five specify the minimum at
$\rho_1=\rho_0$. The set of evolution equations is now given by
\beq
\frac{dm_0^2}{dt} =& \xi_1(0)~~~~~~~~~~
\frac{d\lambda_0}{dt} = \xi_{11}(0)~~~~~~~~~~
\frac{dg_0}{dt} = \xi_{12}(0) \nonumber \\
\frac{d\rho_0}{dt} =& -\frac{\xi_1(\rho_0)}{U_{11}(\rho_0)}~~~~~~~~~~
\frac{dm^2}{dt} = \xi_2(\rho_0) + U_{12}(\rho_0)\; \frac{d\rho_0}{dt}
\nonumber \\
\frac{d\lambda_1}{dt} =& \xi_{11}(\rho_0)
+ U_{111}(\rho_0)\; \frac{d\rho_0}{dt}~~~~~~~~~~
\frac{d\lambda_2}{dt} = \xi_{22}(\rho_0)
+ U_{122}(\rho_0)\; \frac{d\rho_0}{dt}  \nonumber \\
\frac{dg}{dt} =& \xi_{12}(\rho_0)
+ U_{112}(\rho_0)\; \frac{d\rho_0}{dt},
\label{evo8}
\eeq
where $\xi=\partial U_k/\partial t(\rho_1,\rho_2)$ is defined
in eq. (7), with subscripts again denoting partial
derivation in the $\rho$-directions.
The quantities on r.h.s. of the flow
equations (\ref{evo8}) are computed by neglecting terms
with more than four derivatives with
respect to $\rho_1$, $\rho_2$. The remaining partial derivatives
(as for example $U_{1122}$) are expressed in terms of the parameters
defined in eqs. (\ref{var8}) by use of the polynomial approximation
of eq. (\ref{pot8}). This leads to a system of eight coupled
non-linear differential equations
\footnote{The reduced system of three equations
(\ref{DRAX}),(\ref{DLAX}),(\ref{DGAX}) is obtained by the truncation
$\nu_A=\nu_B=\gamma_A=\gamma_B=\gamma_C=0$ and selecting the
couplings $\rho_0$, $\lambda_1$ and $g$.}.

We have solved this system numerically, starting with
a classical potential given by eq. (\ref{POT1})
and
$\mu^2(\Lambda)/\Lambda^2=2.22 \times 10^{-4}$,
$\lambda(\Lambda)=1/90$, $g(\Lambda)=1/10$.
As long as $\lambda_1(k)$ remains larger than about
$10^{-4}$ we see small quantitative but no qualitative
differences as compared to the reduced system
of eqs. (\ref{DRAX}),(\ref{DLAX}),(\ref{DGAX}).
The trajectories
look very similar to the ones of a second order
phase transition. Only if we start with $\kappa(\Lambda)$
so close to the critical value that
$\lambda_1(k)$ runs to values much smaller than
$10^{-4}$ the first order character of the transition is
revealed. We have depicted the phase diagram for small
values of $\lambda_1$ in fig. 2, where we show the
trajectories in the
$(\lambda_1, \kappa)$ plane.
The third relevant
\footnote{We do not distinguish here between relevant and
marginal couplings in the language of statistical mechanics.}
coupling $g$ remains almost constant
in this part of the phase diagram.
When starting the trajectories in the lower right corner of fig. 2
all the remaining ``irrelevant'' parameters
$(m_0^2, \lambda_0, g_0, m^2, \lambda_2)$
have to be set with very high precision equal to the values they take
after the running along the (almost) critical trajectory for
$\lambda_1(k)>10^{-4}$.
Since this first part of the running extends over many orders of
magnitude in the scale $k$ and the critical trajectory is unstable
in the relevant $\kappa$-direction,
care has to be taken to keep track of the critical trajectory
accurately
\footnote{
The inherent numerical uncertainties do not permit to follow the
almost critical trajectories in one sweep from $k=\Lambda$
to $k=0$. Algorithms with a ``returning'' of $\kappa$ at
intermediate $k$ had to be developed. For the
``starting point'' for fig.2 we used values for the
parameters
approximately given by
$m_0^2/k^2=-1.36 \times 10^{-4},
\lambda_0=6.74 \times 10^{-5},
g_0=4.18 \times 10^{-2},
\kappa=\rho_0/k^2=2.09,
m^2/k^2=8.73 \times 10^{-2},
\lambda_1=6.39 \times 10^{-5},
\lambda_2=1.88 \times 10^{-5},
g=4.17 \times 10^{-2}$.}.

The right and lower corner of the phase diagram of fig. 2
corresponds to an effective average potential $U_k$ with only
one minimum. On the other hand, if a trajectory crosses the dashed
line a second minimum develops and the upper left corner of fig. 2
corresponds to a potential with two local minima at $\rho=0$ and
$\rho =\rho_0 \not= 0$. To the left of the dashed-dotted line
the minimum at the origin is deeper than the minimum at
$\rho_0 \not= 0$. The curve B (which asymptotically approaches this
line) corresponds to the phase transition line which separates
the symmetric phase (all trajectories below B) from the phase with
spontaneous symmetry breaking (all trajectories above B).
The SSB phase can be subdivided into region I with only one minimum
of $U_k$ and region II with minima at
$\rho=0$ and $\rho =\rho_0 \not= 0$, the second one being deeper. The
boundary trajectory A separates these regions. Similarly,
in region III the potential has two local minima with the
deeper one at zero. Region III is bounded by the spinodal curve
C. This is the trajectory where $\lambda_1$ vanishes asymptotically
for $k \rightarrow 0$. For trajectories below C and above D the
minimum at $\rho_0 \not= 0$ disappears for some scale
$k_2 > 0$, where $\lambda_1(k_2)=0$. Finally, for trajectories
below the boundary D the location of the minimum $\kappa(k)$
reaches zero for some scale $k_s >0$ with
$\lambda_1(k_2)>0$. The behaviour of all trajectories above A and
below D resembles closely a second order phase transition. The
typical behaviour of the first order transition is only visible
in a narrow range around the critical line in figs. 1 and 2 which
corresponds to the region between the curves A and D. We should
mention
at this point that for trajectories between A and D we have stopped
the running of the couplings at some scale $k_{cnv} >0$
instead of computing the effective potential for
$k \rightarrow 0$. The scale $k_{cnv}$ was chosen large enough
so that $k^2+M^2_1> \frac{1}{4} k^2$, and
$k^2+M^2_2> \frac{1}{4} k^2$ for all
$\rho_1$, $\rho_2$ (cf. eqs. (\ref{MAST}), (\ref{MS})), and small
enough so that the running couplings approached almost constant
values. The potential with several minima therefore corresponds to the
effective average potential $U_{k_{cnv}}$, rather than the
effective potential $U_0$. In other words, the quantum
fluctuations with momenta $q^2 < k^2_{cnv}$ have not been included.
The properties of the phase transition can actually be
understood in terms of the shape of
$U_{k_{cnv}}$. Including the quantum fluctuations with
$q^2 < k^2_{cnv}$
will lead to a convex effective potential $U_0$ \cite{convex}.
The physics related to the ``approach to convexity'' is not
relevant for our purpose and rather obscures the simple
picture presented above.

For $k>k_{cnv}$ we find that the contribution of the
$\varphi_1$ modes to the running of $U_{11}(\rho_1,\rho_2=0,k)$
is very small as compared to the contribution of the
$\varphi_2$ modes, once $\lambda_1(k)$ becomes smaller
than $10^{-4}$. The $\varphi_1$ modes may therefore be
neglected in this region
\footnote{Even though suspected by the smallness of
$\lambda_1$ a similar statement cannot be proven
within the standard perturbation theory since the saddle
point appoximation breaks down for small
$\rho_1$.},
corresponding to an omission of the first
term on the r.h.s. of the flow equation
(\ref{MAST}).
With
\be
M^2_2(\rho_1,\rho_2=0)=U_2(\rho_1)=m^2_0+g_0\rho_1
+\frac{1}{2}\nu_A \rho^2_1
+\frac{1}{6}\gamma_B \rho^3_1
\label{em2} \ee
and neglecting the $k$-dependence of
$m^2_0, g_0, \nu_A, \gamma_B$ we can easily integrate eq.
(\ref{MAST}). The solution for $k \rightarrow 0$
\footnote{Since only the fluctuations of $\varphi_1$ are related
to the approach to convexity we can replace here
$k_{cnv}$ with zero.}
corresponds to a one-loop formula
\beq
U_{11}(\rho_1,0)=& \lambda_0 + \nu_B \rho_1 + \frac{1}{2}
\gamma_C \rho^2_1
\nonumber \\
&-\frac{1}{32 \pi^2} \left( g_0 + \nu_A \rho_1 + \frac{1}{2}
\gamma_B \rho^2_1 \right)^2
\int^{\infty}_0 dx \; x \biggl\{
\frac{1}{(x+U_2(\rho_1))^2}
-\frac{1}{(P_{k_1}(x)+U_2(\rho_1))^2}
 \biggr\} \nonumber \\
&+\frac{1}{32 \pi^2} \left( \nu_A  +
\gamma_B \rho_1 \right)
\int^{\infty}_0 dx \; x \biggl\{
\frac{1}{x+U_2(\rho_1)}
-\frac{1}{P_{k_1}(x)+U_2(\rho_1)}
 \biggr\}.
\label{chris} \eeq
Here $k_1$ may be chosen so that
$\lambda_1(k_1)=10^{-4}$,
with $P_{k_1}=x \left[ 1 - \exp \left( - \frac{x}{k^2_1} \right)
\right]^{-1}$. All couplings on the r.h.s. are taken at the scale
$k_1$, with $\nu_A$, $\nu_B$, $\gamma_B$ and $\gamma_C$
re-expressed in terms of the couplings defined in eqs. (\ref{var8}).
Omitting all couplings except
$m^2_0$, $\lambda_0$ and $g_0$, we recognize the usual
logarithmic behaviour of the Coleman-Weinberg potential
\be
U_{11}(\rho_1,0) \approx  \lambda_0 - \frac{g_0^2}{32 \pi^2}
\left[ \ln \left( \frac{k^2_1+m^2_0+g_0\rho_1}{m^2_0+g_0\rho_1}
\right) -\frac{k^2_1}{k^2_1+m^2_0+g_0\rho_1}
\right].
\label{appro} \ee
Eq. (\ref{chris}) is in agreement with our numerical solution of the
evolution equations and constitutes a quantitative improvement
since no polynomial truncation of $U_k(\rho_1,0)$ is imposed
for $k<k_1$.

In conclusion, a combination of the flow equations for
$\lambda_1(k)>10^{-4}$ with the ``one-loop formula'' (\ref{chris})
for $\lambda_1(k)<10^{-4}$
permits a reliable and quantitatively very precise understanding
of the first order phase transition. Our non-perturbative
method clearly answers the shortcomings of perturbation theory and
firmly
establishes the first order character of the transition for
$\lambda(\Lambda) < \frac{1}{3} g(\Lambda)$. The situation is
analogous to the abelian Higgs model for which a similar
non-perturbative investigation has been performed recently
\cite{LW}. Our method can be easily adapted
to non-vanishing temperature or to the three-dimensional
model relevant for statistical mechanics. In fact, we have
presented our equations already for arbitrary dimensions
$d$ and we emphasize that the usual perturbative infrared
divergences in three dimensions are absent, due to the
presence of an infrared cutoff $\sim k$. For a study
of the temperature dependence of the effective
potential it is sufficient to replace the threshold functions
of eq. (\ref{sdn}) by temperature dependent threshold
functions \cite{TW}, the modification arising from the
discretization of the zero-component of the momentum.
Combining the method
presented here with the study of temperature dependence in ref.
\cite{TW} constitutes an excellent basis for a quantitative
understanding of the high temperature behaviour of the
two-scalar model. This study is presented in detail
in ref. \cite{BTW}.

\section*{Figures}

\renewcommand{\labelenumi}{Fig. \arabic{enumi}}
\begin{enumerate}
\item  
Phase diagram for $\lambda(\Lambda) = 0.1$.
\item  
First order phase diagram in the region of small
$\lambda_1$.
\end{enumerate}

\end{document}